\documentstyle[pra,aps,multicol,epsfig]{revtex}

\input epsf.tex

\newcommand{\be}{\begin{equation}}
\newcommand{\ee}{\end{equation}}
\newcommand{\ba}{\begin{eqnarray}}
\newcommand{\ea}{\end{eqnarray}}
\newcommand{\ban}{\begin{eqnarray*}}
\newcommand{\ean}{\end{eqnarray*}}

\newcommand{\sandwich}[3]{\mbox{$ \langle #1 | #2 | #3 \rangle $}}
\newcommand{\ket}[1]{\mbox{$ | #1 \rangle $}}
\newcommand{\bra}[1]{\mbox{$ \langle #1 | $}}

\newcommand{\si}{\sigma}
\newcommand{\demi}{\frac{1}{2}}
\newcommand{\compl}{\begin{picture}(8,8)\put(0,0){C}\put(3,0.3){\line(0,1){7}}\end{picture}}

\newcommand{\one}{\leavevmode\hbox{\small1\normalsize\kern-.33em1}}

\def\tr{{\rm tr}}

\begin{document}

\title{Bell inequalities and distillability in $N$-quantum-bit systems}
\author{Antonio Ac\'\i n$^1$, Valerio Scarani$^1$ and Michael M. Wolf$^2$}
\address{
$^1$Group of Applied Physics, University of Geneva, 20, rue de
l'Ecole-de-M\'edecine, CH-1211 Geneva 4, Switzerland\\
$^2$Institut f\"{u}r Mathematische Physik, TU Braunschweig,
Mendelssohnstrasse 3, 38106 Braunschweig, Germany}
\date{\today}
\maketitle

\begin{abstract}
The relation between Bell inequalities with two two-outcome
measurements per site and distillability is analyzed in systems of
an arbitrary number of quantum bits. We observe that the violation
of any of these inequalities by a quantum state implies that
pure-state entanglement can be distilled from it. The
corresponding distillation protocol may require that some of the
parties join into several groups. We show that there exists a link
between the amount of the Bell inequality violation and the size
of the groups they have to form for distillation. Thus, a strong
violation is always sufficient for full $N$-partite
distillability. This result also allows for a security proof of
multi-partite quantum key distribution (QKD) protocols.
\end{abstract}

\begin{multicols}{2}

\section{Introduction}

The statistical correlations between the outcomes of experiments
performed by different observers in a composed quantum system can
in general not be reproduced by local variable models (LV) (in the
sense of Einstein-Podolsky-Rosen \cite{EPR}). This impossibility
is shown by proving that quantum correlations violate some
constrains, known as Bell inequalities \cite{Bell}, that any LV
theory satisfies. Thus, a quantum state $\rho$ in  a composite
systems of $N$ parties,
$\compl^{d_1}\otimes\compl^{d_2}\otimes\ldots\otimes\compl^{d_N}$,
where $d_i$ is the dimension of the Hilbert space associated to
party $i$ ($i=1,\ldots,N$), doesn't admit a LV description when it
violates a Bell inequality. It is not difficult to see that
separable states,
\begin{equation}\label{sep}
  \rho=\sum_j p_j
\ket{\psi_1^j}\bra{\psi_1^j}\otimes\ket{\psi_2^j}\bra{\psi_2^j}\otimes
\ldots\otimes\ket{\psi_N^j}\bra{\psi_N^j} ,
\end{equation}
i.e. those states that can be written as a mixture of product pure
states, do not violate any Bell inequality \cite{Werner}. States
that are not separable are called entangled. Entanglement is then
a necessary condition for the violation of a Bell inequality.

The understanding and interpretation of quantum correlations has
notably changed in the last years. Entanglement has turned out to
be a practical resource, since it is the key ingredient for most
of the recent quantum information applications, such as
teleportation \cite{telep} and quantum key distribution
\cite{qkd}. In all these new information processing protocols,
some results that cannot be achieved in Classical Information
Theory become possible by using entangled states. These processes
do not have classical analog because they are based on
entanglement, which is an intrinsic quantum feature. Nevertheless,
it is not clear whether all entangled states are useful for
quantum information tasks.

In systems of two parties, $\compl^{d_1}\otimes\compl^{d_2}$, the
most representative entangled state is
\begin{equation}\label{maxent}
  \ket{\Psi}=\frac{1}{\sqrt d}\sum_{i=1}^d\ket{i}\otimes\ket{i} ,
\end{equation}
where $d=\min(d_1,d_2)$ and $\{\ket{i}\}$ are orthonormal bases in
the two local systems. The state (\ref{maxent}) is the maximally
entangled state of two $d$-dimensional systems, often called {\sl
qudits}. A state $\rho$ is definitely useful for quantum
information applications when out of possibly many copies of it,
the parties are able to distill some amount of maximally entangled
states using only local operations and classical communication
(LOCC). If this is the case, the state $\rho$ is said to be
distillable \cite{dist}. This condition is equivalent to see if
some pure-state entanglement can be extracted from the original
state, since all entangled pure states are distillable
\cite{psdist}. It is known that there are mixed states, called
bound entangled, that are not distillable in spite of being
entangled \cite{bound}. It is an open question whether this type
of states are useful for quantum information. For systems of more
parties the picture is more complicated, and it is not even known
what the fundamental types of pure-state entanglement are
\cite{npart}. However, as we will see, one can extend the notion
of distillability to the multipartite scenario: a quantum state
shared by $N$ parties is $N$-party distillable when out of many
copies of it all the parties can extract by LOCC pure-state
$N$-party entanglement, i.e. a pure state that is bipartite
entangled with respect to any splitting of the parties into two
groups.

Distillability and the violation of Bell inequalities are two
manifestations of entanglement. The first is related to the
usefulness of a state for quantum information processing. On the
other hand, Bell inequality violation demonstrates the inadequacy
of classical LV models. Is it possible to relate these two
concepts? This is the main motivation for the present work: to
search for a connection between Bell inequality violation and
distillability. We consider systems of $N$ quantum bits, or
qubits, and the complete set of Bell correlation inequalities with
two two-outcome measurement per site (see below). We demonstrate
that there exists a link between their violation and
state-distillability.

The structure of the article is the following. In the next section
we describe more precisely our $N$-qubit scenario. We introduce
the family of Bell inequalities we consider and we extend the
concept of distillability to these systems. In Section
\ref{secbip} we establish a first result: if a $N$-qubit state
violates a Bell inequality of this family, it is at least
bipartite distillable. This result is used as a basis for the main
result of the paper, described in Section \ref{secamount}: the
amount of violation is connected to the {\em degree of
distillability} of the state. In particular, if the violation of
an $N$-qubit inequality exceeds the bound $2^{(N-2)/2}$, the state
is fully distillable. These results are valid for the whole family
of inequalities that we study. We move then to consider a subset
of these inequalities that detects truly $N$-qubit entanglement,
and we prove that their violation is also sufficient for $N$-party
distillability (section \ref{secuffink}). Finally, in Section
\ref{secqkd} we discuss the connection of these results with the
security of multi-partite key distribution protocols, and in
Section \ref{secconcl} we summarize our work, relating it to some
existing results for $N$-qubit systems. Three appendices contain
the most technical steps of the demonstrations, that we have
chosen not to include in the main text in order to enhance its
readability and to underscore the physical strength of the
results.

\section{$N$-qubit systems}
\label{secnqub}

In this article we deal with states shared by an arbitrary number
of observers, $N$, such that the dimension of each local Hilbert
space is two (qubits). Let us describe here more precisely the
type of Bell inequalities and distillability protocols that we
consider in these systems.

\subsection{Bell inequalities}

A complete set of Bell correlation inequalities for $N$-qubit
systems was found by Werner and Wolf, and independently Zukowski
and Brukner, in \cite{WW,ZB}. Every local observer,
$i=1,\ldots,N$, can measure two observables, $O^1_i=O_i$ and
$O^2_i=O'_i$, of two outcomes labelled by $\pm 1$. Thus, after
many rounds of measurements, all the parties collect a list of
experimental numbers, and they can construct the corresponding
list of correlated expectation values,
$E(j_1,j_2,\ldots,j_N)=\langle O^{j_1}_1\otimes
O^{j_2}_2\otimes\ldots\otimes O^{j_N}_N\rangle$, where $j_i=1,2$.
The general expression for the Werner-Wolf-Zukowski-Brukner (WWZB)
inequalities is given by a linear combination of the correlation
expectation values,
\begin{equation}\label{complset}
  I_N(\vec c)=\sum_{j_1,\ldots,j_N} c(j_1,\ldots,j_N)
  E(j_1,\ldots,j_N)\leq 1 ,
\end{equation}
where the conditions for the coefficients $\vec c$ can be found in
\cite{WW,ZB}. This set is complete in the following sense. If none
of these inequalities is violated, there exists a LV model for the
list of data $E(j_1,\ldots,j_N)$. If any of these inequalities is
violated, the observed correlations do not admit a LV description.
Thus, this family of inequalities can be thought of as the
generalization of the CHSH inequality \cite{CHSH,Fine},
\begin{equation}\label{CHSHin}
  I_2=\frac{1}{2}(E(1,1)-E(1,2)+E(2,1)+E(2,2))\leq 1
\end{equation}
 to an arbitrary
number of subsystems. Indeed it reduces to the CHSH inequality
when $N=2$.

Now consider a system composed of $N$-qubits and quantum
observables corresponding to von Neumann measurements \cite{WW2}
$\sigma(\hat n)\equiv\hat n\cdot\vec\sigma$, where $\hat n$ is a
normalized three-dimensional real vector and
$\vec\sigma=(\sigma_x,\sigma_y,\sigma_z)$. Thus, any observable is
defined by a real unit vector $\hat n$. It is known that for any
Bell inequality with the corresponding local measuring apparatus,
one can define the so-called Bell operator \cite{BMR}. In our
case, for any of the inequalities (\ref{complset}) and any set of
local measurements $\{\hat n_i,\hat n'_i\}$, with $i=1,\ldots,N$,
we can construct the Bell operator $B=B(\vec c,\{\hat n_i,\hat
n'_i\})$ such that $I_N(\vec c)=\tr(\rho B)$. Then, $\rho$
violates the corresponding Bell inequality if
\begin{equation}\label{bellop}
    \tr(\rho B)>1 .
\end{equation}

The spectral decomposition of all these Bell operators is known
\cite{SG1}, and implies that the maximal violation is always
obtained for Greenberger-Horne-Zeilinger (GHZ) states of $N$
qubits $\ket{\mbox{GHZ}_N}=(\ket{0\ldots 0}+\ket{1\ldots
1})/\sqrt{2}$. Indeed, one can find (i) a local computational
basis $\{\ket{0}_j,\ket{1}_j\}$ for each qubit $j$, (ii) $2^{N-1}$
non-negative numbers $b_k$ and (iii) $2^{N-1}$ parameters
$\theta_k$ such that \ba \label{spectr}
B_N&=&\sum_{k=0}^{2^{N-1}-1}b_k\, \big(Q_{k}^{+}-Q_{k}^{-}\big)\ea
where $Q_{k}^{\sigma}$ are the projectors on the generalized
GHZ-states $\ket{\theta_k^\sigma}$ defined as \ba \label{thbasis}
\ket{\,\theta_k^{\pm}}&=&\frac{1}{\sqrt{2}}
\big(e^{i\theta_k}\ket{k}\pm\ket{\bar{k}}\big)\,. \ea  In these
expressions we label the product states in the computational basis
by $\ket{k}$ with $k\in\{0,1,...,2^N-1\}$, where the
correspondence is given by the binary expansion, i.e.
$\ket{0}=\ket{0...0}$, $\ket{1}=\ket{0...01}$, and so on until
$\ket{2^N-1}=\ket{1...1}$. We also define
$\ket{\bar{k}}=\ket{2^N-1-k}$; written as tensor product,
$\ket{\bar{k}}$ is obtained from $\ket{k}$ by exchanging all the
zeros and ones. The set of the $\ket{\theta_k^{\sigma}}$ with $k
\in\{0,1,...,2^{N-1}-1\}$ and $\sigma=\pm$ is a basis of
eigenstates of $B_N$, that we shall call the {\em theta basis}.
The values of the coefficients $b_k$ and $\theta_k$ depend on the
specific Bell's inequality and the chosen measurements \cite{SG1}.

An important member of this family of inequalities is the
Mermin-Belinskii-Klyshko (MBK) inequality \cite{Mermin,GBP}. Given
a set of measurements $\{\hat n_i,\hat n'_i\}$, the $N$-qubit Bell
operator for these inequalities is defined recursively as
\begin{eqnarray}\label{mermop}
    M_N&=&\frac{1}{2}\big[(\sigma(\hat n_N)+\sigma(\hat n'_N))
    \otimes M_{N-1} \nonumber\\
    &+&  (\sigma(\hat n_N)-\sigma(\hat n'_N))\otimes
    M'_{N-1}\big] ,
\end{eqnarray}
where $M'_n$ is obtained from $M_n$ interchanging $\hat n_i$ and
$\hat n'_i$, and $M_1=\sigma(\hat n_1)$. The maximal quantum
violation of the set of inequalities (\ref{complset}) is obtained
for the MBK one \cite{WW}, with some particular choice of
measurements, and it is equal to $2^{(N-1)/2}$, i.e. quantum
violations of WWZB inequalities are in the range
$(1,2^{(N-1)/2}]$.

\subsection{Distillability}

The concept of state distillability is very often related to the
usefulness of a state for quantum information tasks. In the
bipartite case, a state $\rho$ is distillable when out of possibly
many copies of it the two parties can extract a two-qubit
maximally entangled state, or singlet,
\begin{equation}\label{singlet}
  \ket{\Phi}=\frac{1}{\sqrt 2}(\ket{00}+\ket{11}) ,
\end{equation}
by LOCC. It is not completely evident how to extend this
definition to a multi-partite scenario. In this work we will use
the following generalization: a quantum state shared by $N$
parties is $N$-partite distillable when it is possible to distill
states like (\ref{singlet}) between any pairs of parties using
LOCC. This is equivalent to demand that any truly $N$-partite
entangled state, in particular an $N$-qubit GHZ state, can be
obtained by LOCC. Indeed, once all the parties are connected by
singlets, one of them can prepare locally any of these states and
send it to the rest by teleportation.  On the other hand if the
parties share an $N$-partite pure entangled state, there exists
local projections such that $N-2$ qubits project the remaining two
parties into a bipartite entangled pure state \cite{PR}, which is
always distillable to a state like (\ref{singlet}).

In the multipartite case the situation is subtler than in the
bipartite one. Consider a state $\rho$ which is not $N$-partite
distillable. It may happen that if some of the parties join into
several groups (or establish quantum channels between them), the
state becomes distillable (see \cite{DC}). The original state is
now shared by $L<N$ parties, and it is $L$-partite distillable.
However it is important to stress that some of the parties can
extract singlets without using any global quantum operation
between them. They already had some distillable entanglement that
was hidden and that can be extracted by joining some of the
parties. We have two extreme cases: (i) the parties can perform
all the operations locally, and then the state is $N$-partite
distillable as was defined above, or (ii) they have to join into
two groups, and then the state is said to be bipartite distillable
\cite{DC}. All the other cases are between these two possibilities
and, of course, there also non-distillable states. According to
this classification, we will estimate the degree of distillability
in an $N$-qubit state by means of the minimal size of the groups
the parties have to create in order to distill pure-state
entanglement between them.

\section{Violation of Bell inequalities implies bipartite distillability}
\label{secbip}

One of the present authors has proven some time ago \cite{Acin}
that for a {\em specific} Bell operator $B_N$, namely the MBK
operator with some given settings, all the states that violate the
corresponding Bell inequality are bipartite distillable, although
the partition may not be such that one of the parties is a single
qubit \cite{Dur}. Refs. \cite{Acin,Dur} are the only two studies
of the link between violation of Bell and distillability in
$N$-qubit systems before this one \cite{zuk}. As a first step, we
provide the generalization of the result of Ref. \cite{Acin} for
an arbitrary inequality of the WWZB family.

{\em Theorem 1} Consider an $N$-qubit state $\rho$. If there
exists a Bell operator $B_N$ in the WWZB family such that the
corresponding inequality is violated, that is such that
$\mbox{Tr}(\rho B_N)>1$, then $\rho$ is bipartite distillable.

{\em Proof:} The proof goes along two steps.

{\em First step.} If $\rho$ is such that $\mbox{Tr}(\rho B_N)>1$
for a Bell operator with two measurement per qubit, then it has
been proven in Ref. \cite{WW} that there exists at least one
partition of the $N$ qubits into two disjoint groups, $A$ and
$A^{c}$, such that the partial transpose \cite{parttr} of $\rho$,
$\rho^{T_A}$, has a negative eigenvalue (the state is said to be
NPT). This  condition is necessary for distillability
\cite{bound}, but it is conjectured not to be sufficient
\cite{NPTBES}, except for $\compl^2\otimes\compl^d$ systems. Now,
from the state $\rho$, we form the state \ba
\rho_D\,=\,\sum_{k=0}^{2^{N-1}-1} \sum_{\sigma=\pm}
\lambda_k^{\sigma}\,Q_k^{\sigma}\,,&\mbox{ with }&
\lambda_k^{\sigma}\equiv\mbox{Tr}(\rho\,Q_k^{\sigma})
\label{rhod}\ea by keeping only the terms that are diagonal in the
theta basis associated to $B_N$. Note that we do not claim that
there is an LOCC operation such that $\rho \mapsto \rho_D$. By
construction, $\mbox{Tr}(\rho B_N)=\mbox{Tr}(\rho_D B_N)$:
$\rho_D$ violates the same inequality as $\rho$. As a consequence
of the result in Ref. \cite{WW}, we know that there exists at
least one bipartite splitting $A-A^c$ of the qubits such that
$\rho_D^{T_A}$ has a negative eigenvalue. As stressed several
times, this is not in general a guarantee that $\rho_D$ is
distillable. However, it turns out that here NPT {\em is} a
sufficient condition for distillability.

\begin{center}
\begin{figure}
\epsfxsize=8cm \epsfbox{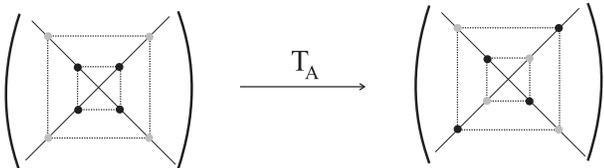} \caption{Schematic
representation of the effect of the partial transposition on the
matrix $\rho_D$, written in the product basis.} \label{figdiag}
\end{figure}
\end{center}

To see this, we write the matrix $\rho_D$ in the product basis.
This gives \ba
\rho_D&=&\sum_{k=0}^{2^{N-1}-1}\,\Big[\frac{\lambda_k^++\lambda_k^-}{2}\,
\big(\ket{k}\bra{k}+\ket{\bar{k}}\bra{\bar{k}}\big)\,+\nonumber\\
&&+ \frac{\lambda_k^+-\lambda_k^-}{2}\,
\big(e^{i\theta_k}\ket{k}\bra{\bar{k}}+\mbox{h.c.}\big)\Big]\,.
\ea This means that there are non-zero elements only in the two
main diagonals of the matrix. It is easy to get convinced that the
partial transposition with respect to any partition $A-A^c$ will
preserve this structure (fig. \ref{figdiag}). As an example, take
$N=5$ qubits, $A=\{2,3\}$. Then \ba
\ket{0\underline{01}11}\bra{1\underline{10}00}
&\stackrel{T_A}{\longrightarrow}&\ket{0\underline{10}11}\bra{1\underline{01}00}\,:
\ea an element of the anti-diagonal is sent onto another element
of the anti-diagonal. Of course, the elements of the main diagonal
remain unchanged.

This being the structure of $\rho_D^{T_A}$, the negativity of some
eigenvalue can be studied by looking at each $2\times 2$ bloc \ba
M(k)&=&\left(\begin{array}{cc} \big(\rho_D^{T_A}\big)_{kk} &
\big(\rho_D^{T_A}\big)_{k\bar{k}}\\
\big(\rho_D^{T_A}\big)_{\bar{k}k} &
\big(\rho_D^{T_A}\big)_{\bar{k}\bar{k}}
\end{array}
\right)\,. \ea for $k=0,...,2^{N-1}-1$. Since $\rho_D$ is NPT for
some bipartite splitting, there exist at least two values of $k$,
say $K$ and $K'$, and a bipartite splitting $A-A^c$ such that: (i)
$\ket{K'}\bra{\bar{K'}}$ is sent onto $\ket{K}\bra{\bar{K}}$ by
the partial transposition $T_A$, and (ii) the determinant of the
$2\times 2$ block $M(K)$ is negative: \ba
(\lambda_K^++\lambda_K^-)^2 -
(\lambda_{K'}^+-\lambda_{K'}^-)^2&<&0\,.\ea Now it is not
difficult to see that the $N$-qubit state $\rho_D$ is bipartite
distillable. According to the partition $A-A^c$, the state can be
locally projected into the subspace ${\cal{H}}(K,K')$ spanned by
$\ket{K}$, $\ket{\bar{K}}$, $\ket{K'}$ and $\ket{\bar{K'}}$
\cite{noteloc}. This subspace is isomorphic to
$\compl^2\otimes\compl^2$, and one can re-label
$\ket{K}=\ket{00}$, $\ket{\bar{K}}=\ket{11}$, $\ket{K'}=\ket{01}$
and $\ket{\bar{K'}}=\ket{10}$. The projected two-qubit state
satisfies NPT, then it is distillable \cite{horo}.

This is the end of the first step of the proof. The only problem
left is that the transformation $\rho\longrightarrow\rho_D$ may
well be impossible with LOCC \cite{noteloc}. We have not a final
proof for this impossibility. Note however that the theta basis in
which we should diagonalize $\rho$ is defined by $2^{N-1}$ highly
non-local parameters, the phases $\theta_k$, whose value is
determined by the details of the observable $B_N$ \cite{SG1}. We
cannot get easily rid of the $\theta_k$ by using the freedom left
in the construction, that is by a redefinition of local phases
$\ket{0}_j\rightarrow e^{i\phi_{j}^0}\ket{0}_j$ and
$\ket{1}_j\rightarrow e^{i\phi_{j}^1}\ket{1}_j$, because there are
only $2N$ such phases. As we said in the main text, instead of
looking for an hypothetical LOCC protocol leading from $\rho$ to
$\rho_D$ defined in (\ref{rhod}), we take another approach.

{\em Second step.} In the previous step, we have identified the
subspace in which to project the state, locally for the partition
$A-A^c$. It is the subspace ${\cal{H}}(K,K')$ spanned by
$\ket{K}$, $\ket{\bar{K}}$, $\ket{K'}$ and $\ket{\bar{K'}}$, or
alternatively, by $\ket{\theta_K^{+}}$, $\ket{\theta_K^{-}}$,
$\ket{\theta_{K'}^{+}}$ and $\ket{\theta_{K'}^{-}}$. We begin by
applying a local phase redefinition $U$ that erases the phases
$\theta_K$ and $\theta_{K'}$: thus $U\ket{\theta_{K,K'}^{\pm}}=
\ket{\psi_{K,K'}^{\pm}}$ where the $\psi$'s are the GHZ states
without phases \ba\label{ghzbas}
\ket{\psi_k^{\pm}}&=&\frac{1}{\sqrt{2}}
\big(\ket{k}\pm\ket{\bar{k}}\big)\,. \ea This $U$ of course does
not erase all the other phases $\theta_k$, but this is not a
problem.

D\"{u}r and Cirac \cite{DC} have shown that any $N$-qubit state
$\rho$ can be brought by LOCC to a state diagonal in a GHZ-basis
like (\ref{ghzbas}), \ba \label{depprot}
\rho'_{D}=\sum_{k=0}^{2^{N-1}-1} \sum_{\sigma=\pm}
\mu_k^{\si}\,P_k^{\sigma}\,,&\mbox{ with }&
\mu_k^{\sigma}\equiv\mbox{Tr}(\rho\,P_k^{\sigma}) \ea where
$P_k^{\sigma}$ is the projector on the GHZ state
$\ket{\psi_k^{\pm}}$. All the diagonal terms,
$\bra{\psi_k^\pm}\rho\ket{\psi_k^\pm}$, are kept unchanged, while
the rest of terms go to zero. Thus in our case we can bring $U\rho
U^{\dagger}$ onto \ba \rho'_D&=& \sum_{\sigma=\pm}
\left(\lambda_{K'}^{\si}\,P_{K'}^{\sigma}+
\lambda_K^{\si}\,P_K^{\sigma}\right)\,+\, \sum_{k\neq K,K'}
\sum_{\sigma=\pm} \mu_k^{\si}\,P_k^{\sigma}\,, \ea just using
local operation on each sub-system. Note that for $K$ and $K'$ the
$\lambda_k^{\sigma}$ are indeed the same that appear in the
construction (\ref{rhod}) of $\rho_D$. Obviously, when written in
the product basis, $\rho'_D$ has exactly the same structure as
$\rho_D$, that is, non-zero elements only in the two main
diagonals. Contrary to what happened for $\rho_D$, we do not know
if $\rho'_D$ violates the original Bell's inequality. However this
is not important here, we simply have to apply the same procedure
that we followed for $\rho_D$: take the partition $A-A^c$ that
brings $\ket{K'}\bra{\bar{K'}}$ onto $\ket{K}\bra{\bar{K}}$; then
by construction the determinant of $M'(K)$ built from
${\rho'}^{T_A}_D$ is the same as the determinant of $M(K)$ built
from ${\rho}^{T_A}_D$. Thus $A-A^c$ can locally project $\rho'_D$
into ${\cal{H}}(K,K')$, and the resulting two-qubit state will
satisfy NPT and will thus be distillable. This concludes the
proof. $\Box$

In summary, the way of distilling a singlet from a state $\rho$
that violates a Bell inequality is: (a) determine on the paper the
$2\times 2$ subspace ${\cal{H}}(K,K')$ in which to project and the
corresponding partition; (b) erase locally the phases $\theta_K$
and $\theta_{K'}$ and apply the D\"{u}r-Cirac protocol, and
finally (c) project onto ${\cal{H}}(K,K')$. In the proof, we used
three known results: the spectral decomposition for Bell operators
with two measurements per qubit \cite{SG1}, the fact that any
$\rho$ that violates one of these Bell inequalities is NPT for at
least one partition \cite{WW}, and the depolarization protocol of
\cite{DC}. The new insight is provided by the peculiar structure
of the matrices $\rho_D$ and $\rho'_D$ that makes NPT a sufficient
condition for distillability.

\section{The amount of violation and the degree of distillability}
\label{secamount}

\subsection{Main result}

In this section we prove the main result of the article: there
exists a link between the amount of Bell violation and the degree
of state distillability. We have just shown that if a state
violates any of the WWZB inequalities (\ref{complset}), then it is
bipartite distillable. As it has been mentioned above, the range
of quantum violations, $(1,2^{(N-1)/2}]$, is quite broad,
specially for a large number of qubits. This suggests that a finer
classification of state distillability properties can be done
depending on the amount of Bell violation. This is the scope of
this section. Let us start by proving the following

{\em Lemma 1:} Consider an $N$-qubit state $\rho_N$ that violates
an inequality of (\ref{complset}), with Bell operator $B_N$, by an
amount $\beta_N$, i.e.
\begin{equation}\label{nlcond}
    \tr(\rho_N B_N)=\beta_N>1 .
\end{equation}
Then, it is possible to obtain by LOCC a new state $\rho_{N-1}$ of
$N-1$ qubits violating another inequality of (\ref{complset}),
with Bell operator $B_{N-1}$, by an amount
$\beta_{N-1}\geq\beta_N/\sqrt 2$.

{\em Proof:} It was shown in \cite{WW} that the CHSH is the
elementary inequality for the whole set (\ref{complset}). This
means that for $N$ qubits any of these inequalities can be written
as
\begin{eqnarray}\label{chshel}
    B_N&=&\frac{1}{2}\big[(\sigma(\hat n_N)+\sigma(\hat n'_N))
    \otimes B^+_{N-1} \nonumber\\
    &+&  (\sigma(\hat n_N)-\sigma(\hat n'_N))\otimes
    B^-_{N-1}\big] ,
\end{eqnarray}
where $B^\pm_{N-1}$ are WWZB Bell operators of $N-1$ qubits --- of
course, the special relation that $B^{-}_N$ is obtained from
$B^{+}_N$ by interchanging $\hat n_i$ and $\hat n'_i$ holds only
for the MBK inequality. Using local unitary operations, the $N$th
qubit can put $\hat{n}_N$ and $\hat{n}_N'$ in the $xy$ plane,
their bisectrix being the $x$ axis. Denote by $2\delta$ the angle
between the two vectors, $0\leq\delta\leq\pi/2$. Then, we have
that the state $\rho_N$ satisfies
\begin{eqnarray}\label{cond}
    \tr(\rho_N B_N)&=&\cos\delta\,\tr(\rho_N\,\sigma_x
    \otimes B^+_{N-1}) \nonumber\\
    &+& \sin\delta\,\tr(\rho_N\,\sigma_y\otimes B^-_{N-1})=\beta_N .
\end{eqnarray}
Suppose now that $\tr(\rho_N\,\sigma_x\otimes B^+_{N-1})\geq
\tr(\rho_N\,\sigma_y\otimes B^-_{N-1})$ (of course a similar
demonstration is possible for the other case). Then it follows
from (\ref{cond}) that
\begin{equation}\label{cond2}
    \tr(\rho_N\,\sigma_x\otimes
    B^+_{N-1})\geq\frac{\beta_N}{\cos\delta+\sin\delta}\geq
    \frac{\beta_N}{\sqrt 2} .
\end{equation}
If we use the spectral decomposition
$\sigma_x=\ket{+}\bra{+}-\ket{-}\bra{-}$, and we denote by $\tilde
B^+_{N-1}\equiv -B^+_{N-1}$, which is a new Bell operator of $N-1$
qubits, we have
\begin{equation}\label{cond3}
    \tr(\bra{+}\rho_N\ket{+}B^+_{N-1})+
    \tr(\bra{-}\rho_N\ket{-}\tilde B^+_{N-1})\geq\frac{\beta_N}{\sqrt 2} .
\end{equation}
Define the normalized states
$\rho_\pm\equiv\bra{\pm}\rho_N\ket{\pm}/p_\pm$ of $N-1$ qubits,
where $p_\pm\equiv\tr(\bra{\pm}\rho_N\ket{\pm})$. The physical
meaning of these states is the following: if the $N$ qubits start
with state $\rho_N$ and party $N$ measures $\sigma_x$, the rest of
the qubits are projected into $\rho_\pm$ with probability $p_\pm$.
Again without loss of generality, consider the case in which
$\tr(\rho_+B^+_{N-1})\geq\tr(\rho_-\tilde B^+_{N-1})$. Then from
(\ref{cond3}) it is easy to see that
\begin{equation}\label{cond4}
    \tr(\rho_+B^+_{N-1})=\beta_{N-1}\geq\frac{\beta_N}{\sqrt 2} .
\end{equation}
Thus, starting from $\rho_N$ that has a Bell violation equal to
$\beta_N$, any qubit can locally project with some probability the
other $N-1$ qubits into a new state $\rho_{N-1}$ that violates a
new inequality by an amount of at least $\beta_N/\sqrt 2$. $\Box$

We note that $\beta_{N-1}\geq \beta_N/\sqrt{2}$ can always be
obtained with non-zero probability. In the case where $p_+$ (or
$p_-$) is zero, $\rho_N$ is a product state containing a
$\sigma_x$ eigenstate on the relevant tensor factor. However,
since $\bra{+}\sigma_y\ket{+}=\bra{-}\sigma_y\ket{-}=0$, Eq.
(\ref{cond}) tells us that in this situation $\beta_N=\beta_{N-1}$
right from the beginning, such that no measurement is required at
all.

Of course, it is likely that some of the inequalities used in the
derivation of this lemma are not tight. However they cannot be
improved if we do not have more information about the specific
state or Bell operator. We can now combine this lemma with the
result shown in Section \ref{secbip} for proving the following

{\em Theorem 2:} Consider an $N$ qubit state $\rho_N$ violating
one of the WWZB inequalities by an amount $\beta$ such that
\begin{equation}\label{viol}
    1<2^{\frac{N-p}{2}}<\beta\leq 2^{\frac{N-p+1}{2}} .
\end{equation}
Then pure-state entanglement can be distilled if the parties can
join into groups of at most $p-1$ qubits.

{\em Proof:} Using the lemma seen above, any qubit can perform a
local projection such that the amount of Bell violation is
decreased by a factor $\sqrt 2$. In the worst case, after $N-p$ of
these local projections, the rest of $p$ qubits share a state
$\rho_p$ having a Bell violation of $1<\beta_p\leq\sqrt 2$. A new
local projection is not possible since it might imply that the
resulting state is not entangled. At this point, and since
$\rho_p$ is still non-local, we can use the result of Section
\ref{secbip}: the state $\rho_p$ is bipartite distillable. Thus,
these $p$ qubits can distill pure-state entanglement between them
if they can join into groups of at most $p-1$ parties. $\Box$

\begin{center}
\begin{figure}
\epsfxsize=8cm \epsfbox{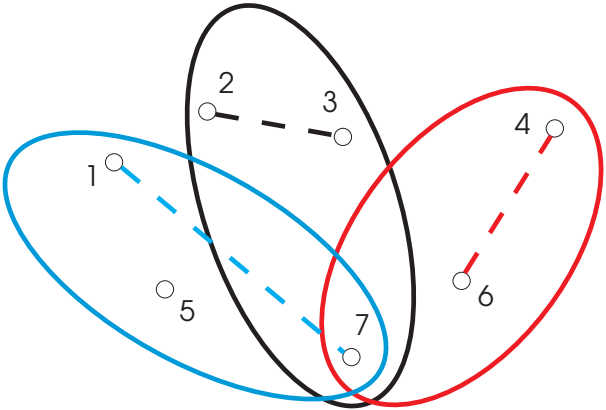}
\bigskip\bigskip \caption{The
figure shows the case of a seven-qubit state having a Bell
violation of $4 \sqrt 2$. After local projections by four parties,
the other three are left with a state violating a WWZB inequality.
Then, if two of them share a quantum channel, they can distill
pure-state entanglement. For the blue partition, for instance,
parties 2, 3, 4 and 6 perform the local projection in such a way
that 1, 5 and 7 end with a three-qubit state violating a WWZB
inequality. Now, they can distill entanglement if (at most) two of
them share a quantum channel, say parties 1 and 7. The parties run
this protocol for all the different groups as in the figure, and
at the end party 7 can prepare any $N$-qubit pure state locally
and use teleportation for sending the corresponding qubit to the
rest of parties. Thus, any $N$-qubit state can be prepared between
all the parties using several copies of the initial state,
provided that quantum channels are established between some of
them.} \label{distfig}
\end{figure}
\end{center}

This theorem gives the searched link between Bell violation and
the degree of state distillability. As in any distillation
scenario we have at our disposal many copies of the state
$\rho_N$. Thus, the parties can use the different copies for
connecting all of them. The amount of Bell violation bounds the
size of the groups they have to form in any of these distillation
processes. It gives an estimation of the $L$-partite
distillability of the state ($2\leq L\leq N$), or in other words,
the number of quantum channels to be established between the
parties for distillation (see figure \ref{distfig}). If we focus
now on $N$-partite distillability, we have

{\em Corollary 1:} If an $N$-qubit state violates any of the WWZB
inequalities by an amount $\beta_N>2^{(N-2)/2}$, then it is
$N$-partite distillable.

{\em Proof:} It follows easily from the previous theorem. All the
parties but 1 and $l$, with $l=2,\ldots,n$, perform the local
projection. Then the two qubits $l$ and 1 end, with some
probability, with a state $\rho_{1l}$ violating the CHSH
inequality. This two-qubit state is entangled and then it is
distillable \cite{horo}. In this way, the first party shares
singlets with all the others, so the initial state is $N$-party
distillable. $\Box$

\subsection{Some comments}

Another possible manifestation of entanglement is the negativity
of the partial transposition \cite{parttr}. If a state $\rho_N$
has a non-positive partial transposition (the state is NPT) with
respect to some bipartite splitting of the parties, then it is not
separable for this splitting. It is also known that if the state
is distillable, then it is NPT \cite{bound}. The relation between
Bell violation and partial transposition in $N$-qubit systems was
analyzed in \cite{WW3} for the MBK case. Since distillability is
sufficient for NPT, our results are a generalization of the ones
in \cite{WW3} to the whole set of WWZB inequalities. Indeed,
defining as $\beta_{max}$ the maximal Bell violation for an
$N$-qubit state $\rho$, if $\beta_{max}>2^{(N-p)/2}$, every subset
containing $p$ parties has at least one NPT partition. In a
similar way as in \cite{WW3}, one can consider a partition of the
$N$ qubits into $p$ nonempty and disjoint subsets
$\alpha_1,\ldots,\alpha_p$ and the collection ${\mathcal P}$ of
all unions of these sets with the empty set. The set ${\mathcal
P}$ has $2^p$ elements. Then, if $\forall\alpha\in{\mathcal P}$ we
have $\rho^{T_\alpha}>0$, then $\beta_{max}\leq 2^{(N-p)/2}$. In
particular, if $\beta_{max}>2^{(N-2)/2}$, all the partitions are
NPT (as it should be, since the state is fully distillable).

As it has been already mentioned, it may happen that for a
particular situation the bounds presented here are not good.
However, they cannot be improved: a better estimation of the
distillability properties of an $N$-qubit state is not possible if
we only know the amount of Bell violation. The clearest example is
a $N$-qubit state of the form $\ket{\mbox{GHZ}_{N-1}}\ket{0}$.
This state violates the MBK inequality for $N$ qubits by a quite
large amount, $2^{(N-2)/2}$, even if it is clearly not $N$-partite
distillable --- even worse, one of the qubits is not entangled at
all. Similar examples for GHZ states of $N-p+1$ qubits in
$N$-qubit systems show that the bounds given in Theorem 1 are
indeed tight. In the next Section, we show that additional
knowledge about the {\em meaning} of a given inequality leads to
an improvement of these bounds.

\subsection{Amount of violation and weight of the GHZ state}

Some further insight on the meaning of the main result above can
be gained by noting that a state must have a large overlap with
the GHZ state in order to violate an inequality of the WWZB family
by a large amount. This is not astonishing, and can be quantified.

Given the $N$-qubit state $\rho_N$, one can always redefine local
bases (or apply local unitary operations) in order to maximize \ba
r&=&\sandwich{\mbox{GHZ}_N}{\,\rho_N\,}{\mbox{GHZ}_N}\,. \ea It
can be proven (see Appendix A) that for all Bell operator $B$ in
the WWZB family, normalized so that the LV limit is at 1, it holds
\ba\label{bounds}
    \tr(\rho_N\,B)& \leq&\beta(r)\,=\,2^{\frac{N-1}{2}}\,
    \sqrt{r^2+\frac{(1-r)^2}{2^{N-1}-1}}\,.
\ea A necessary condition to detect $N$-qubit entanglement is
therefore $\beta(r)>2^{(N-2)/2}$. Let us see what this condition
implies when the number of qubits is varied.

For {\em two qubits}, the condition $\beta(r)>1$ is fulfilled for
$r>1/2$; but $r$ is the weight of a Bell state in $\rho$, and it
is known that $r>1/2$ is a sufficient condition for the state to
be entangled. Consequently, for two qubits this bound simply says
that if a state violates a Bell inequality, then it is entangled
(and thence distillable).

For {\em three qubits}, the condition $\beta(r)>\sqrt{2}$ is
fulfilled for $r>(1+\sqrt 3)/4\sim 0.683$. In Appendix B, we show
that this is sufficient but not necessary for full distillability,
by giving an explicit protocol. As expected, the violation of a
Bell inequality by a large amount is a sufficient, but not a
necessary condition for full distillability. This is in perfect
analogy with the two-qubit case, where there are distillable (that
is, entangled) states that do not violate any Bell inequality.
This is also in agreement with the facts that states of the form
$\cos\alpha\ket{0...0}+\sin\alpha\ket{1...1}$, for an arbitrary
number of qubits, never give a large violation (if any) when
$\alpha$ is small enough \cite{zb2}; however, such states are
distillable to $\ket{\mbox{GHZ}_N}$ by filtering and classical
communication.

In the limit of a {\em large number of qubits}, we have
$\beta(r)\sim 2^{(N-1)/2}\,r$. This means that for a violation
$\beta_N>2^{(N-p)/2}$, $p\geq 2$, as in Theorem 1, the overlap $r$
with the GHZ state must be larger than $2^{(1-p)/2}$. In
particular, for the violation implying $N$-qubit entanglement
($p=2$) one must have $r>1/\sqrt 2\sim 0.71$.

\section{$N$-party entanglement and $N$-party distillability}
\label{secuffink}

In this Section we focus more specifically on full distillability.
We have proven in Corollary 1 that if an $N$-qubit state violates
any of the WWZB inequalities (\ref{complset}) by an amount
$\beta_N>2^{(N-2)/2}$, then it is $N$-partite distillable. This is
a very general result, and because of this generality it cannot be
improved: only $\beta_N>2^{(N-2)/2}$ guarantees truly $N$-qubit
entanglement for the MBK inequalities \cite{GBP}, then {\em a
fortiori} for the {\em whole} set of WWZB inequalities; and we
have just proven that the $N$-qubit entanglement detected by this
criterion is fully distillable.

Recently, other inequalities have been constructed whose violation
guarantees $N$-qubit entanglement \cite{svetl,Uffink}. However,
the amount of violation is smaller. Specifically, for these
inequalities $\beta_N=2^{(N-2)/2}$ is the maximum amount of
violation allowed by QM, and the criterion for $N$-qubit
entanglement reads $\beta_N>2^{(N-3)/2}$. Thus, the general
criterion of Corollary 1 is not fulfilled. It seems, however,
reasonable to conjecture that the $N$-qubit entanglement detected
by these specific inequalities is also fully distillable. We are
going to prove that this is indeed the case. But first, we must
introduce the inequalities under consideration.

\subsection{Uffink's inequality}

In Ref. \cite{Uffink}, Uffink discussed a quadratic inequality
that detects $N$-qubit entanglement. The experimental data for it
are the same as for the WWZB family: each party can perform two
von Neumann measurements, $\{\hat n_i,\hat n'_i\}$. The settings
are chosen in order to maximize
\begin{eqnarray}\label{uffdef}
    U_{N}(\rho)&=&\sqrt{\tr(\rho M_N)^2+\tr(\rho M'_N)^2}\,,
\end{eqnarray}
where $M_N$ and $M'_N$ are the MBK operators defined above. The LV
limit is $U_{LV}=\sqrt{2}$, since in LV, as well as for quantum
product states, the average value of both $M_N$ and $M_N'$ can
reach 1. The QM bound is found to be
$U_{N}(\rho)=2^{\frac{N-1}{2}}$, and if
\begin{eqnarray}\label{uffineq}
    U_{N}(\rho)&>&2^{\frac{N-2}{2}}\
\end{eqnarray}
then $\rho$ exhibits $N$-qubit entanglement. Of course, since the
LV limit is set to $\sqrt{2}$ instead of being set to 1, this
corresponds to a violation $\beta_N>2^{\frac{N-3}{2}}$. At first
sight, Uffink's inequality looks fundamentally different from the
WWZB set of inequalities, since these are linear constraints while
Uffink's parameter $U_N(\rho)$ involve squaring correlation
coefficients. However, using the basic optimization
$\sqrt{x^2+y^2}=\max_{\gamma}\big(\cos\gamma\,x+\sin\gamma
\,y\big)$, one can rewrite (\ref{uffdef}) as
$U_{N}(\rho)=\max_{\gamma}\tr(\rho\,U_{N,\gamma})$ where we have
defined the linear operator
\begin{eqnarray}\label{uffop}
    U_{N,\gamma}&\equiv& \cos\gamma \,M_N\,+\, \sin\gamma \, M'_N\,.
\end{eqnarray} Thus, Uffink's quadratic inequality
turns out to be a compact way of writing a set of linear
inequalities, which are satisfied if the inequalities belonging to
the WWZB family are satisfied. In particular, there exists
$\gamma$ such that $U_{N}(\rho)= \tr(\rho\,U_{N,\gamma})$. In a
geometrical picture, Uffink's parameter $U_{N}(\rho)$ defines a
circle in the plane given by $x=\langle M_N\rangle=\tr(\rho M_N)$
and $y=\langle M'_N\rangle=\tr(\rho M'_N)$. The $U_{N,\gamma}$
define all the tangents to this circle: if a point lies outside
the circle, it also lies beyond some tangent to the circle. In
summary, Uffink's result reads: if there exist some settings and
an angle $\gamma$ such that $\tr(\rho\,U_{N,\gamma})>2^{(N-2)/2}$,
then the state $\rho$ has $N$-qubit entanglement.

Note that the ``generalized Svetlichny inequalities" discussed in
Ref. \cite{svetl}, that also detect $N$-partite entanglement, are
included in the discussion of the Uffink's inequality. In fact,
the operators defining these last inequalities are
$S_N=U_{N,0}=M_N$ for $N$ even and $S_N=U_{N,\pi/4}$ for $N$ odd;
and the condition for $N$-partite entanglement is exactly
$\tr(\rho\,S_N)>2^{(N-2)/2}$.

Before showing that (\ref{uffineq}) implies full distillability,
we want to motivate our interest in Uffink's inequality by showing
that this inequality is indeed {\em stronger} than the MBK
inequality. On the one hand, it is evident from (\ref{uffdef})
that $\tr(\rho\,M_N)>2^{(N-2)/2}$ implies $U_N(\rho)>2^{(N-2)/2}$,
even for the same settings. On the other hand, we can exhibit
states for which $U_N(\rho)>2^{(N-2)/2}$ but $\tr(\rho\,M_N)\leq
2^{(N-2)/2}$. For instance, consider the three-qubit state
$\cos\alpha\ket{000}+\sin\alpha\ket{W}$, $\alpha\in [0,\pi/2]$,
where $\ket{W}=\big(\ket{110}+\ket{101}+\ket{011}\big)/\sqrt{3}$.
We found numerically that the MBK inequality is violated for
$\alpha>\pi/0.63$, while the Uffink's inequality is violated for
$\alpha>\pi/8$. It is not astonishing that Uffink's inequality is
stronger than those previously reported, because it allows
optimization not only on the settings $\{\hat n_i,\hat n'_i\}$,
but also on a non-trivial parameter $\gamma$.

In conclusion, we have an inequality whose absolute violation is
weaker than the MBK, but which is a better detector of $N$-partite
entanglement. We turn now to the proof that the violation of this
inequality also implies full distillability.


\subsection{Distillability from Uffink's inequality}

To show full distillability from the violation of Uffink's
inequality, we begin by applying similar techniques as above.
Indeed, the two MBK operators appearing in (\ref{uffop}) can be
written as
\begin{eqnarray}\label{mermdesc}
    M_N&=&\frac{1}{2}\left((\sigma(\hat n_N)+\sigma(\hat n'_N))
    \otimes M_{N-1}\right. \nonumber\\
    &+& \left. (\sigma(\hat n_N)-\sigma(\hat n'_N))\otimes
    M'_{N-1}\right) ,
\end{eqnarray}
and similarly for $M'_N$. Now, consider that we start with an
$N$-qubit state $\rho_N$ satisfying
$\tr(\rho_N\,U_{N,\gamma})>2^{(N-2)/2}$ for some $\gamma$. If
qubit $N$ applies the same measurement as in the previous section,
the other $N-1$ qubits can be projected with some probability into
a state of $N-1$ qubits violating an inequality $U_{N-1,\gamma'}$.
Thus, the different parties apply this projection until the point
where three of them end with a three-qubit state satisfying
$\tr(\rho_3\,U_{3,\tilde\gamma})>\sqrt{2}$.

So we reduce the proof that the violation of the Uffink inequality
implies full distillability to the simplest case, the one
involving three qubits. Now, using the techniques introduced in
\cite{SG1}, it can be shown (Appendix C) that a state satisfying
$\tr(\rho_3\,U_{3,\tilde\gamma})> \sqrt{2}$ is necessarily such
that \ba
\sandwich{\mbox{GHZ}}{\,\rho_3\,}{\mbox{GHZ}}&\geq&0.628\,. \ea In
this case, the distillation protocol of Appendix B can be applied:
a three-qubit state violating the Uffink inequality is three-party
distillable. Going back, this means that if
$\tr(\rho_N\,U_{N,\gamma})> 2^{(N-2)/2}$, GHZ states of three
qubits can be distilled between any group of three parties, which
means that $\rho_N$ is $N$-partite distillable. This concludes the
proof.

It is interesting to note that in some cases fully distillability
can be associated to a small Bell violation. Indeed consider a
systems of three qubits and the Svetlichny inequality
$S_3=U_{3,\frac{\pi}{4}}$ \cite{svetl}. The value attained by LV
models is exactly the bound above which we have $3$-party
entanglement, i.e. $\sqrt 2$. Thus, in this case an infinitesimal
Bell violation $\tr(\rho_3\,S_3) =\sqrt{2}+\epsilon$ is sufficient
for {\em full} distillability \cite{notesv}.

\section{Bell inequalities and the security of QKD protocols}
\label{secqkd}

The existence of a link between Bell inequality violation and the
security of QKD protocols was first noticed in \cite{crypto}.
There it was shown that the violation of the CHSH inequality is a
necessary and sufficient condition for the security of the BB84
protocol \cite{BB}, under the assumption of individual attacks and
using privacy amplification \cite{ie3}.

This connection was later extended \cite{SGcr} to the following
multi-partite QKD schemes: a sender encodes a key into $N-1$
qubits shared between $N-1$ observers in such a way that all of
them must cooperate in order to retrieve the key. The quantum
version of this protocol, also known as secret sharing, uses the
correlations in a GHZ state of $N$ qubits (see \cite{secr} for
details). The main result of \cite{SGcr} was to show that the
mutual information between the sender and the authorized partners
(all the receivers) is greater than the one between the sender and
the unauthorized partners (the eavesdropper and the dishonest
receivers) if and only if the authorized partners can violate the
MBK inequality by an amount greater than $2^{(N-2)/2}$. It has to
be emphasized that this security criterion is based on the very
plausible fact that the difference between the mutual information
for honest and dishonest parties should allow for some kind of
classical privacy amplification protocol. However the existence of
this protocol is at the present unknown.

The region for security coincides with the Bell violation
sufficient for $N$-party distillability. Thus, if the parties
share a state with the sufficient amount of Bell violation, they
can use the quantum distillation protocol shown here for
distilling a GHZ state and then run the quantum secret sharing
protocol \cite{note2}. In this way Eve is disentangled from the
honest parties, and the protocol is secure. Note that this is a
{\em quantum privacy amplification} protocol, while, as we have
just stressed, the existence of a classical protocol for these
schemes in this security region remains an open question.

Thus the results of the present paper prove that (i) the
"plausible" security criterion put forward in \cite{SGcr} is
definitely a security criterion, at least for quantum privacy
amplification; and (ii) the criterion is extended to the violation
of any of the WWZB inequalities by $\beta_N>2^{(N-2)/2}$, and to
all the inequalities allowing for $N$-partite distillability like
Uffink's. Further investigation in this direction is still
required, but our results strengthen the interpretation of Bell
inequalities as security indicators for quantum cryptography
schemes.

\section{Discussion}
\label{secconcl}

Bell inequalities are usually presented as a way for testing
Quantum Mechanics vs LV theories. Then, its importance is normally
related to the insight they give in our interpretation of the
quantum world. Recently, due to the new quantum information
applications, the current understanding of quantum correlations
has significantly changed. Nowadays it is important to detect when
the correlations in a quantum state are useful for quantum
information tasks, i.e. they are distillable. In this work we have
shown that Bell inequalities can also be useful for this role.
Indeed, since present knowledge on multi-particle entanglement is
far from being exhaustive, they provide a powerful tool for
understanding distillability properties of $N$-qubit states.

What does a multi-particle Bell violation exactly mean from the
point of view of non-locality? It is clear that Bell violation
implies some form of non-locality if want to describe the
correlations within classical probability theory (i.e. within a
hidden variable model). However it has been stressed recently that
even larger violations of an $N$-party Bell inequality may be not
sufficient for truly $N$-party non-locality \cite{svetl}. As an
example, consider a three-qubit state saturating the maximal
violation of the MBK inequality $M_3$. An hybrid non-local model
where two of the parties have non-local correlations and the third
is separated can reach the same value. This leads to the search
for inequalities that detect full $N$-party non-locality. It turns
out that these inequalities are equal to the usual MBK one for the
case of an even number of particles, and to the so-called
Svetlichny inequality (that is, with the notation of this article,
$U_{N,\pi/4}$) for odd $N$ (see \cite{svetl} for details). From
our results it follows that in both cases, full non-locality
implies full distillability.

Full distillability seems to be a quite strong entanglement
criterion for multi-particle states. Indeed the set of local
operations assisted with classical communication becomes less
powerful when the number of parties increases. From the point of
view of Bell inequalities, a quite large violation is needed for
$N$-party distillability. Taking into account that the range of
Bell violations is quite broad for large $N$, it is likely that to
demand $N$-party distillability is much more stringent than to
demand Bell violation. It is interesting to analyze the result
given in \cite{Dur} in view of the ideas presented in this
article. In \cite{Dur} D\"ur constructs a multi-qubit state
violating the MBK inequality that has all local partial
transpositions positive, which means that the parties acting alone
cannot distill entanglement at all. His result fits quite well
into our picture: a strong violation is never reached by this
state, since this would imply $N$-partite distillability.

In conclusion, in this paper we have shown the existence of a link
between Bell violation and state distillability in $N$-qubit
systems: one can estimate the degree of distillability of an
$N$-qubit state from the amount of its Bell violation (or
non-locality). In this way Bell inequalities provide information
about the usefulness of the state for quantum information
applications. Indeed, a strong Bell violation is sufficient for
the security of multi-particle QKD protocols.

\bigskip
\bigskip
\bigskip

We thank Nicolas Gisin, Serge Massar and Wolfgang D\"ur for
discussions. We acknowledge financial support from the Swiss NCCR
"Quantum Photonics" and OFES, within the project EQUIP
(IST-1999-11053) and the DFG.

\section*{Appendix A}

The spectral decomposition of any Bell operator in the WWZB family
is of the form (\ref{spectr}). The $b_k$ are non-negative and we
suppose $b_0$ to be the largest eigenvalue, since we can always
relabel the local bases in such a way that the largest eigenvalue
corresponds to $Q_{0}^{+}$. Moreover by a local unitary
transformation we absorb the phase $\theta_0$, so
$Q_0^+=P_{0}^{+}$ is the projector onto $\ket{\mbox{GHZ}_N}$. If
the Bell operator is normalized so that the LV is set to 1, then
the eigenvalues satisfy the constraint given in Eq. (25) of
\cite{WW}: \ba \sum_{k=0}^{2^{N-1}-1}{b_k}^2&\leq&
2^{N-1}\label{constr1} \ea with equality for the MBK operators.
For each $N$-qubit state $\rho_N$, we have \ba \tr(\rho_N B_N)&=&
\sum_{k=0}^{2^{N-1}-1}b_k\,(\lambda_{k}^{+}-\lambda_{k}^{-}) \ea
with $\lambda_{k}^{\sigma}$ given in (\ref{rhod}). We suppose that
$\lambda_{0}^{+}=r=\sandwich{\mbox{GHZ}_N}{\rho_N}{\mbox{GHZ}_N}$.
is the maximum of the $\lambda_{k}^{\sigma}$. By keeping only the
positive terms, $\lambda_k\equiv\lambda_k^+$, and normalizing the
probabilities so that \ba \sum_{k=1}^{2^{N-1}-1}\lambda_k&=&1-r\,.
\label{constr2}\ea we obtain the upper bound \ba \tr(\rho_N
B_N)&\leq&
\beta(r)\,=\,\sup\left(b_0\,r\,+\,\sum_{k=1}^{2^{N-1}-1}b_k
\lambda_k\right)\ea where the supremum is taken over the $b_k$
compatible with (\ref{constr1}) and the $\lambda_k$ satisfying
(\ref{constr2}). Using Lagrange multipliers, it turns out that for
fixed $b_0$ the extremum is reached when
\begin{equation}
  \lambda_k=\frac{1-r}{2^{N-1}-1} \quad\quad
  b_k=\sqrt{\frac{2^{N-1}-b_0^2}{2^{N-1}-1}},
\end{equation}
for all $k=1,...,2^{N-1}-1$. Writing $b_0=2^{N-1}\cos\eta$, we
obtain for $\beta(r)$ the expression \ba \beta(r)&=&\max_{\eta}
\left(r\cos\eta+(1-r) \frac{\sin\eta}{\sqrt{2^{N-1}-1}}
\right)2^{\frac{N-1}{2}}\,. \ea Using
$\max_{\eta}\big(\cos\eta\,x+\sin\eta \,y\big)=\sqrt{x^2+y^2}$ we
get (\ref{bounds}).

Just one comment to point out a common mistake when dealing with
these Bell operators. We have said that the bound $\tr(\rho_N
B)\leq \beta(r)$ is rough in general. One might expect however
that the bound is exact for the MBK operator and the state
\begin{equation}
  \rho_N(r)=r P_{0}^{+}+\frac{1-r}{2^{N-1}-1}
  \sum_{k=1}^{2^{N-1}-1}P_k^{+}.
\end{equation}
This guess would be correct if (\ref{constr2}), with equality,
were the unique constraint on the eigenvalues of the MBK operator;
but this is not the case \cite{SG1}. Therefore it may happen that
the set of $b_k$ that optimize $\beta(r)$ is not a set of possible
eigenvalues. This remark already applies to the case of three
qubits: as discussed in the text, $\beta(r)=\sqrt{2}$ for
$r=r_3\simeq 0.683$. However numerically one can verify that
$\rho_3(r_3)$ does {\em not} violate the Mermin inequality; the
family of states $\rho_3(r)$ starts to violate the Mermin
inequality only at $r\sim 0.687$.

\section*{Appendix B}

Here we present a simple protocol for full distillability for
three qubits. There is no claim of efficiency, even less of
optimality, for such a protocol. Let us consider the basis of GHZ
states (\ref{ghzbas}) for three qubits. As usual, we shall write
$\ket{\psi_0^{+}}=\ket{\mbox{GHZ}_3}$. One can locally depolarize
any state $\rho$ onto a state which is diagonal in the GHZ basis
keeping the diagonal terms as in (\ref{depprot}). This is the {\em
first step} of the distillation protocol:
\begin{equation}
  \rho\longrightarrow\rho'_{D}=\sum_{k=0}^{2^{N-1}-1} \sum_{\sigma=\pm}
  \mu_k^{\si}\,P_k^{\sigma}
\end{equation}
With further local operations, we can arrange (just for
definiteness) that $\mu_0^{+}$ is the maximum of the
$\mu_k^{\sigma}$.

In the {\em second step}, one of the three parties, say Charlie,
measures his own qubit in the $\sigma_x$ basis and communicates
the result $s^C_x=\pm 1$ to the other two parties Alice and Bob.
This way, Alice and Bob share several copies of each of the two
two-qubit conditional states $\rho_{AB}(s^C_x=\pm 1)$. The idea
now is rather trivial: if at least one of these states is
entangled, then Alice and Bob can distill a singlet out of many
copies of it. Without loss of generality, we can concentrate on
$\rho_{AB}\equiv \rho_{AB}(s^C_x=+ 1)$. In the computational
basis, this state
reads \ba \rho_{AB}&=&\left(\begin{array}{cccc} a &&& c\\
&\demi-a& d\\
&d &\demi-a\\
c &&& a
\end{array}\right) \ea
with \ba
  a&\equiv&\demi\sum_{\sigma}(\mu_0^{\sigma}+\mu_3^{\sigma}),
  \nonumber\\
  c&\equiv&\demi\sum_{\sigma}\sigma(\mu_0^{\sigma}+\mu_3^{\sigma}),
  \nonumber\\
  d&\equiv&\demi\sum_{\sigma}\sigma(\mu_1^{\sigma}+\mu_2^{\sigma}).
\ea A necessary and sufficient condition for distillability is
that $(\rho_{AB})^{T_A}$ has at least a negative eigenvalue. In
the present case, due the form of the matrix, this is a very
simple condition to write. Without loss of generality, we can
suppose that the negative eigenvalue is in the block
\begin{equation}
  M=\left(\begin{array}{cc}\demi-a& c\\
    c &\demi-a\\ \end{array}\right).
\end{equation}
This block must have a positive eigenvalue since the trace is
non-negative, so the necessary and sufficient condition for
distillability is simply $\mbox{det}M<0$. After some algebra,
defining $p^{\pm}\equiv \mu_0^{\pm}+\mu_3^{\pm}$, this condition
reads explicitly \ba p^{+}\,+\, p^{-}\,-\,2p^{+}p^{-}&>&\demi\,.
\label{conddist} \ea We don't need to study the domain of validity
of this condition in very sharp detail. Actually, for our purpose
we simply have to notice that if $\mu_0^{+}>1/2$ then
(\ref{conddist}) is satisfied. But
$\mu_0^{+}=\sandwich{\mbox{GHZ}_3}{\,\rho\,}{\mbox{GHZ}_3}$, thus
whenever \ba
\sandwich{\mbox{GHZ}_3}{\,\rho\,}{\mbox{GHZ}_3}&>&\demi
\label{conddist2} \ea Alice and Bob can distill a singlet if they
collaborate with Charlie. But (\ref{conddist2}) is actually
symmetric in the roles of the three parties; then if
(\ref{conddist2}) holds, any pair of parties can distill a
singlet. This is sufficient for full distillability.

\section*{Appendix C}

We present here the detailed proof of the fact that the violation
of the three-qubit Uffink inequality implies full distillability.
This proof uses tools from the spectral decomposition of WWZB Bell
operators. We begin by applying the results of Ref. \cite{SG1} to
the three-qubit operators $M_3$ and $U_{3,\gamma}$.

\subsection{Spectral decomposition of $M_3$}

The settings $\{\hat n_i,\hat n'_i\}$ that define the Bell
operator are supposed to lie in the $(x,y)$ plane for each qubit
$i=1,2,3$ --- this can always be achieved by local unitary
operations on the state. Defining \ba \hat n_i&=&
\cos\alpha_i\,\hat x\,+\,\sin\alpha_i\,\hat y \ea and a similar
definition for $\hat n'_i$, the settings are parametrized by the
angles $\underline{\alpha}=\{\alpha_i,\alpha'_i\}$. We shall also
define $\delta_i=\alpha_i-\alpha'_i$.

For all three qubits, let $\ket{0}$ and $\ket{1}$ be the two
eigenvectors of $\si_z$. To write the operators in a compact way,
for all $k\in\{0,1,2,3\}$, let $\vec{\sigma}(k)$ be the operator
acting as the Pauli matrices in the subspace spanned by
$\ket{"+z"}=\ket{0\,k_2\,k_3}$ and $\ket{"-z"}=\ket{1\,\bar
k_2\,\bar k_3}$, where $(k_2,k_3)$ is the binary expression of $k$
and $\bar k_j=1-k_j$ for $j=2,3$. It follows from (\ref{spectr})
that \ba M_3&=&\bigoplus_{k}b_{k}\,
\hat{n}_{k}\cdot \vec{\sigma}(k)\label{m3}\\
M_3'&=&\bigoplus_{k}b_{k}\, \hat{n}\,'_{k}\cdot
\vec{\sigma}(k)\label{m3p} \ea with \ba
\hat{n}_{k}&=&\cos\theta_{k}\,\hat{x}\,+ \,
\sin\theta_{k}\,\hat{y} \ea and similarly for $\hat{n}\,'_{k}$.
The expressions (\ref{m3}) and (\ref{m3p}) imply in particular the
fact, not stressed explicitly in \cite{SG1}, that $M_3$ and $M_3'$
have the same eigenvalues. The way of obtaining the eigenvalues
$b_k$ and the parameters $\theta_{k}$ from the settings
$\underline{\alpha}$ has been discussed in \cite{SG1}: one has \ba
b_k e^{i\theta_{k}}\,\equiv\, f_{k}(\underline{\alpha})&=&
e^{i(\alpha_1'+\beta_2+\beta_3)}
+ e^{i(\alpha_1+\beta_2'+\beta_3)}\nonumber\\
&&+e^{i(\alpha_1+\beta_2+\beta_3')} -
e^{i(\alpha_1'+\beta_2'+\beta_3')} \label{fomega}\ea with
$\beta_j=k_j\,\alpha_j$. The explicit form of $b_k$ is not very
elegant, but will be needed in what follows: \ba b_k&=&\left[
\Big(\prod_{i}\cos\delta_i\Big)^2 + \Big(\prod_{i}k_i\sin\delta_i
+ \sum_{i}k_i\sin\delta_i \Big)^2 \right]^{\frac{1}{4}} \ea where
$i=1,2,3$ and $k_1=1$.

\subsection{The Uffink operator $U_{3,\gamma}$}

The Uffink operator $U_{3,\gamma}$ reads \ba
U_{3,\gamma}&=&\cos\gamma M_3+\sin\gamma M_3'
\,=\,\bigoplus_{k}u_{k}\, \hat{m}_{k}\cdot \vec{\sigma}(k)\ea
where $\hat{m}_{k}$ is the unit vector along the direction
$\cos\gamma\,\hat{n}_{k}+\sin\gamma\,\hat{n}\,'_{k}$ and where the
eigenvalues are given by \ba u_{k}&=&b_k\,\sqrt{1+\sin
2\gamma\,(\hat{n}_{k}\cdot\hat{n}\,'_{k})}\,. \label{evuffink}\ea
Note that $(\hat{n}_{k}\cdot\hat{n}\,'_{k})=
\cos(\theta_{k}-\theta'_{k})$. Now, it follows from (\ref{fomega})
that \ba m_{k}(\underline{\alpha})\, =\,
f_{k}(\underline{\alpha})\,
f_{k}(\underline{\alpha}')^{*}&=&|b_{k}|^2 \,
e^{i(\theta_{k}-\theta_{k'})} \ea where as usual
$\underline{\alpha}'$ means exchanging the settings $\alpha_i
\leftrightarrow \alpha'_i$. By comparison with (\ref{evuffink}) we
find \ba {u_{k}}^2&=& |m_{k}(\underline{\alpha})|+\sin 2\gamma
\mbox{Re}\big(m_{k} (\underline{\alpha})\big)\nonumber\\ &=&
{b_{k}}^2\,+\,\sin 2\gamma\prod_{i}\cos\delta_i \,.\ea

Here we can apply the same estimate as in Appendix B. Note that
for the Uffink operator the bound (\ref{constr1}) does not hold in
general, since the LV limit is not set to 1. But we have just to
replace that constraint by \ba \sum_{k}{u_{k}}^2&=&4\,\big(1+\sin
2\gamma\,\cos\delta_1\cos\delta_2\cos\delta_3\big) \ea since for
the MBK operators $M_3$ it holds $\sum_{k} {b_{k}}^2=4$. Apart
from that, the calculation with the Lagrange multipliers is the
same. Without loss of generality, we can take $u_0$ to be the
highest eigenvalue of $U_{3,\gamma}$, and with local unitary
operations we can choose $\ket{\mbox{GHZ}_3}=
(\ket{000}+\ket{111})\sqrt{2}$ as the associated eigenvector.
Writing $\underline{\delta}=\{\delta_1, \delta_2,\delta_3\}$, we
obtain \ba \beta_{\gamma}(r)&=&\sup_{\underline{\delta}}\,\left[
u_{0}(\underline{\delta})\,r + \bar{u}(\underline{\delta}) (1-r)
\right] \ea with
$r=\sandwich{\mbox{GHZ}_3}{\,\rho\,}{\mbox{GHZ}_3}$ and
\begin{equation}
  \bar{u}(\underline{\delta})\equiv\left(
\frac{\sum_{k}{u_{k}}^2-{u_{0}}^2}{3}\right)^{\frac{1}{2}} .
\end{equation}
The optimization over $\underline{\delta}$ can be done
numerically. One finds there are some settings
$\underline{\delta}$ and some values $\gamma$ for which one can
get $\beta_{\gamma}(r)>\sqrt{2}$ only if $r\geq r_U\sim 0.628$. In
conclusion: a necessary condition to violate the Uffink inequality
for three qubits is that \ba
\sandwich{\mbox{GHZ}_3}{\rho}{\mbox{GHZ}_3}&\geq&0.628\,. \ea Thus
in particular (\ref{conddist2}) is fulfilled, and we can distill
singlets between any two parties using the protocol described in
Appendix C.

Finally, we want to mention that in this case
\begin{equation}
  \rho_3(r)=r\,P_0^++ \frac{1-r}{3}\sum_{k\neq 0}P_k^{+}
\end{equation}
gives $U_3(\rho_3(r))>\sqrt{2}$ for $r\geq 0.628$. Comparing with
the remark that concludes Appendix A, we see that the same state
gives $\tr(\rho_3(r) M_3)>\sqrt{2}$ only for $r>0.687$. This is
another manifestation of the strength of the Uffink inequality.

\end{multicols}


\begin{thebibliography}{10}
\bibitem{notesv}
Note that this does not exclude that if a violation of this
inequality is observed, one can also obtain a violation of one of
the inequalities in (\ref{complset}) large enough for fully
distillability. However we have some numerical evidence suggesting
that this is not the case.
\bibitem{note2}
Note that once singlets between the sender and the other $N-1$
parties are prepared, they can establish pairwise perfectly
correlated keys and then run classical secret sharing protocols.
\end{thebibliography}
\end{document}